\documentclass[english,twocolumn,preprintnumbers,amsmath,amssymb,showpacs]{revtex4}
\usepackage[T1]{fontenc}
\usepackage[latin1]{inputenc}
\usepackage{graphicx}

\makeatletter

\usepackage{dcolumn}
\usepackage{bm}
\usepackage{times}

\usepackage{multirow}

\usepackage{babel}
\makeatother
\begin{document}

\title{Globally controlled fault tolerant quantum computation}

% Force line breaks with \\

\author{Joseph Fitzsimons }

\email{joe.fitzsimons@materials.ox.ac.uk}

\affiliation{Department of Materials, Oxford University, Oxford, UK}

\author{Jason Twamley}

\affiliation{Centre for Quantum Computer Technology, Macquarie University, Sydney,
NSW 2109, Australia}

\begin{abstract}
We describe a method to execute globally controlled quantum information
processing which admits a fault tolerant quantum error correction
scheme. Our scheme nominally uses three species of addressable two-level
systems which are arranged in a one dimensional array in a specific
periodic arrangement. We show that the scheme possesses a fault tolerant
error threshold. 
\end{abstract}

\maketitle
\begin{figure}
\begin{centering}\setlength{\unitlength}{1cm} \begin{picture}(7,5)
\put(-1.5,0){\includegraphics[width=10cm]{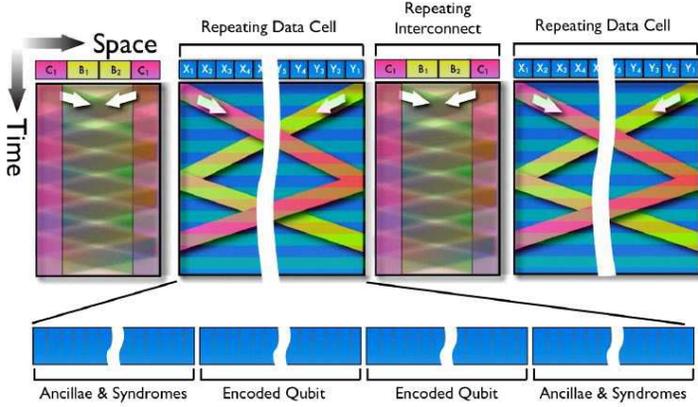}} \end{picture} \par\end{centering}
\caption{
%Mirror symmetric encoding of two logical qubits and associated syndromes in the $A$-species computational block, e.g. for instance the Shor 3-bit code and 2 syndromes. We also indicate the mirror symmetric propagation of quantum information within the $A$-blocks under repeated global applications of ${\cal S}_{A}=\overline{H}_{A}\cdot\overline{CZ}_{A}$ (shaded horizontal bars). 
Spatial arrangement of $A$-species {\em computational blocks} (in blue), and $B$-{\em interconnect blocks} (in yellow), and $C$-{\em reset cells} (pink). 
(a) Using global pulses all three species can be decoupled
and $A,\;B$-blocks mirror cycled. $C$-cells can also be globally reset. (b)  mirror symmetric spatial arrangement of two encoded qubits and their associated error correction sydromes and ancillae qubits.
\label{fig:fig1}}
\end{figure}

\indent Global control provides a novel way of quantum computation
which should greatly reduce the complexity of classical control technology
required in a medium-to-large scale quantum processor. A globally
controlled quantum computer architecture typically only permits one
to apply quantum gates homogeneously on large subsets of the processor
and one is not allowed to target gates on individual qubits within
the processor. A number of designs have appeared in the literature
but so far their usefulness has been hampered by the lack of any design
which also incorporates globally controlled quantum error correction
executed in a fault tolerant manner. In this note we describe a 1D
scheme for such fault-tolerant computation which only includes three addressable qubit species
arranged in a self-similar one dimensional pattern.
 \vspace{.25cm}\\
\textit{\bf Background:-} The study of globally controlled architectures
began with \cite{Lloyd1993}, which used a three species spin chain
arranged in a periodic linear array. In \cite{Benjamin2000}, a two
species 1D design was developed where an {}``always-on'' interaction
was modulated by homogenous local unitaries (HLUs). Models where one
can homogeneously modulate the inter-chain couplings were presented
\cite{Benjamin2001,Levy2002}. Both \cite{Dodd2002} and \cite{Jane2003},
examine the simulation power of a quantum system with always-on interactions
and modulated HLUs. More recently, a number of globally controlled
schemes for 1D quantum computation have been discovered displaying
various levels of sophistication of construction and control \cite{Janzing2005,Ivanyos2005,Raussendorf2005a,Volbrecht2006,Raussendorf2005b,Fitzsimons2006}.
Most of these globally controlled schemes can be cast into two main
categories, (A) those that use special {}``software labels'' (the
{\em control unit}, in \cite{Benjamin2000}), which move via global
pulses within the processor and whose purpose is to effectively localise
an applied global pulse to a local region \cite{Benjamin2000,Benjamin2002,Volbrecht2006}.
The other main category (B), is where one uses {}``hardware labels''
to trigger the conversion of globally applied pulses to qubits within
the device. In \cite{Raussendorf2005b} this is achieved via a change
in global parity of an evolving delocalised qubit pattern upon impacting
with the physical ends of the chain, while in \cite{Fitzsimons2006},
control is achieved by manipulating the delocalised qubit pattern
when it also impacts an end of the chain. One faces a number of challenges
in developing a fault tolerant quantum error correct scheme in a 1D
globally controlled design with nearest-neighbour (n-n) interactions.
From \cite{Aharonov1999,Gottesman2000,Kempe2005}, to implement concatenation
of QEC one at least requires (1) fully parallel execution of quantum
computation, (2) one must ensure that errors do not proliferate, i.e.
one round of computation and error correction is successful in reducing
the overall error rate, (3) methods to remove entropy from the system,
and (4) the error rates do not drastically increase with the concatenation
level. Obviously the restriction to n-n models and the associated
increased error rates due to the shuttling of qubits around to execute
long range gates will prove detrimental to the performance of fault
tolerant quantum error correction but a number of works have now shown
that FTQEC is still possible \cite{Aharonov1999,Gottesman2000,Fowler2004,Svore2005,Szkopek2006,Svore2006}.
We will restrict ourselves below to category (B) designs where one
has a hardware trigger to manipulate the delocalised qubits. Previously
a number of works have examined possible FTQEC schemes for category
(A) designs \cite{Bririd2003,Kay2005,Kay2007}, but there one has
difficulty in correcting for errors in the special {}``software label''
itself using only global control.
 \vspace{.25cm}\\
\textit{\bf Outline:-} We follow \cite{Fitzsimons2006}, where one uses
an always-on Ising ($ZZ$), interaction and a single qubit Hadamard
HLU to construct a global operation ${\cal S}=\overline{H}\cdot\overline{CZ}$,
as a mirror iterate and via edge operations, one fashions universal
quantum computation. Below we shall assume that the spin chain is
subject to temporal errors (which we assume to be independent Pauli
errors \cite{Footnote1}), and we construct a hierarchy of logical meta-qubit encodings, such
that on the highest level we effectively will have a meta-Ising model
where the associated meta-$ZZ$, meta-$\overline{H}$ and meta-$\overline{CZ}$,
operations possess greatly reduced error rates. We then can implement
quantum computation on this meta-cellular automaton using either of
the approaches of \cite{Raussendorf2005b,Fitzsimons2006}. We will
argue that the resulting model possesses a fault tolerant threshold,
however we will here not give specific estimates for the magnitude
of the threshold, limiting ourselves to providing a proof of existence.
 \vspace{.25cm}\\
\textit{\bf Model:-} Our model is based on \cite{Fitzsimons2006}, and
requires at least two addressable spin species. Our scheme is most
clearly illustrated with three separately addressable species arranged
in a specific linear arrangement all coupled via an {}``always on''
$ZZ$ Ising interaction. However the scheme also works with just two
species though in a more complicated fashion. We consider seperate
{\em computational blocks}, made up of $N_{Comp}$ cells of either
species $A$ or $B$ where $N_{Comp}$ is chosen to accommodate the
encoding of two logical qubits via some chosen quantum error correction
scheme and their associated syndrome bits and ancillae in a mirror symmetric spatial
arrangement (Fig.\ref{fig:fig1}). The species $A$ blocks will be
used to store logical qubits, with the species $B$ blocks acting
as a divider between species $A$ encoded blocks. From \cite{Fitzsimons2006},
we know that given the capability of performing edge operations on
the $A$-species {\em computational blocks}, we can perform universal
quantum computation and, in particular, we can execute any given quantum
error correction scheme on the encoded logical qubits. From \cite{Kempe2005},
it is better to use a quantum error correction scheme where one considers
an error operation occurring in the encoded qubit state, record the
associated syndromes in ancilla bits and then apply the recovery operation
directly on the encoded logical qubits. The alternative, decoding
and then recovery, will yield worse thresholds as the decoded qubit
is unprotected for a short period. Thus, within the {\em computational
block}, we can execute the encoding, syndrome recording and error
recovery with universal quantum computation. All that remains is to
reset the syndrome bits ready for the next error correction cycle.
Thus to execute one cycle of quantum error correction on a {\em
computational block} we need to be able to execute edge operations
and also reset syndrome bits. To do this in a fault tolerant manner
one must be be able to perform these operations in parallel and have
a way of performing all the steps required quantum error correction
and the effective ($ZZ$) Ising interaction at any higher concatenation
(or meta), level.

 %\vspace{.25cm}\\
\textit{\bf Base Concatenation Level:-} We first address the steps necessary
at the base (or zero), concatenation level. We arrange the species-$A$
{\em computational blocks} in a periodic arrangement on a linear
chain separated by {\em linking chains}. These linking chains are
the crucial elements of our scheme and are shown schematically in
Fig.\ref{fig:fig1}. These are made up of mirror symmetric arrangements
of single $C$-species spins on either side of two $B$-species sites.
The $C$-type species are differentiated apart from the $A$ and $B$-species
in that they possess a spin-non conserving transition which can be
triggered via an appropriate global pulse sequence to reset the $C$-species
to the $|0\rangle_{C}$ state irrespective of its initial state. 
The
$C$-species provides a local entropy sink for the syndrome bits in
the neighboring {\em computational blocks}, and are used to impliment
local rotations on the end of the neighbouring subchains. The wires
of $B$-species joining $C$-spins can be globally addressed and,
for the most part, act to ferry data between $C$ and $A$ blocks.
We will later use qubits encoded with increasing concatenation levels
in the $B$-wire interconnects to inter-connect meta-qubits on different
concatenation levels. We now show how $C$-species spins can be exploited
to (i) allow the execution of edge operations on the $A$ and $B$-blocks
necessary for universal computation within those blocks and (ii) how
the syndrome bits within the $A$-blocks can reset and then reused
for another round of error correction. To achieve (i) we arrange decoupling
pulses to decouple the three species. This can be achieved by stroboscopically
applying periodic $X$ gates to the $A$ and $B$ species subchains.
We then we reset the $C$ cells (via a seperate global pulse), and
then arrange for a controlled-phase operation between $C$ (now in
the $|0\rangle_{C}$ state), and the neighbouring $A$ and $B$ cells
by halting the $B-C$ and $A-C$ decoupling for a short period. As
the $B-C$ and $A-C$ decoupling are controled seperately it is possible
to effect different controlled phase operations on the end spins of
chains of species $A$ and $B$. For the scheme of \cite{Fitzsimons2006},
all we require for universal computation is the capability of performing
one-qubit phase rotations on the ends of the qubit chain (this coupled
with homogenous unitaries on the $A$- and $B$-blocks are sufficient
to execute any single qubit unitary; and edge operations combined
with the decoupling of the edge sites from the interior of the block
are sufficient to give two-qubit gates). With the $C$-spins adjacent
to the $A$- and $B$-blocks now in the state $|0\rangle_{C}$, the
$A-C$ and $B-C$ Ising couplings, when not decoupled, give $\exp(iJ_{A-C}Z_{A}Z_{C})|\psi_{A}0_{C}\rangle\rightarrow\exp(iJ_{A-C}Z_{A})|\psi_{A}0_{C}\rangle$
and $\exp(iJ_{B-C}Z_{B}Z_{C})|\psi_{B}0_{C}\rangle\rightarrow\exp(iJ_{B-C}Z_{B})|\psi_{B}0_{C}\rangle$
respectively. Thus by decoupling the blocks for a short time we can
execute a phase rotation on the $A$- and $B$-block edge cells. To
achieve (ii) we arrange that the $A$-block syndromes are positioned
in cells adjacent to the $C-$spins. By halting decoupling for an
appropriate period it is possible to generate controlled-$Z$ gates
between the spins of species $C$ and the adjacent qubits. This, combined
with global pulses on each species, is sufficient to construct a SWAP
gate (using the standard triple CNOT construction), between adjacent interfacial $A-C$ and $B-C$ cells.
First, we use the $A-C$ SWAP gate to move the syndrome qubit onto
the $C$-spin, and then use the $B-C$ SWAP to place the syndrome on the adjacent $B$-spin. We then decouple and execute one mirror cycle of the
combined $B$-wires, to move this cell towards the opposite $C$-block. The syndrome is then swapped from the $B$-spin to the $C$-spin. Once the syndrome qubits are localised on the $C$-spins we follow with
a global erasure pulse of all $C$-blocks. Following this all of the
syndrome bits are on the $C$-spins but now reset to zero. We then
reverse the $B$-wire transport and SWAP them back into the $A$-species
computational block. 
%Both (i) and (ii) are schematically illustrated in Fig. \ref{fig3}.
\begin{widetext}
\begin{figure*}
\begin{centering}\setlength{\unitlength}{1cm} \begin{picture}(5,5)
\put(-6,-.3){\includegraphics[height=6.5cm,width=16cm]{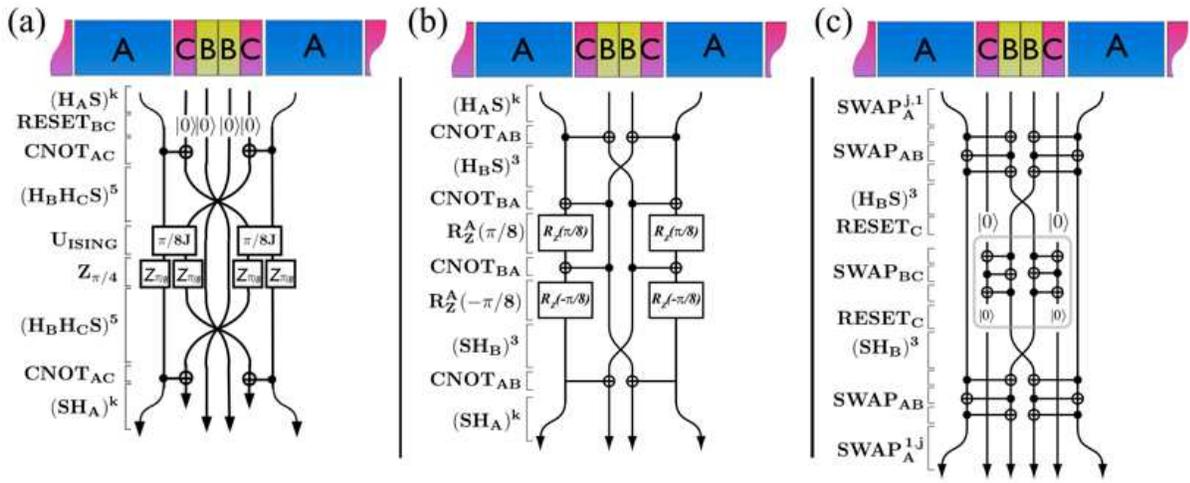}} 
 \end{picture} \par\end{centering}
\caption{Illustration of of gates between the $j^{th}$ physical qubits of the encoded $A-$block qubits, showing a (a) base level concatenation CZ gate between interconnect-seperated $A-$blocks, (b) same at level-$L$ concatenation using $CNOT_{A,C}$ gates. (c) Level-$L$ ancillae reset procedure (see text).
\label{fig:fig2}\vspace{-.7cm}}
\end{figure*}
\end{widetext}
 Thus we have shown how to execute sucessive rounds of quantum error
correction on encoded qubits held in all the $A$-species computational
blocks. However we must also be able to simulate the logical CPHASE
gate between adjacent logical qubits. This is done in two rounds the
first of which is to perform CPHASEs between the pair of logical qubits
encoded symmetrically in each $A$-species computational block. In
many quantum error correction codes executing a CPHASE on the encoded
logical qubit is performed by a transversal CPHASE on each physical
qubit. Since we can perform universal quantum computation in each
$A$-block executing such CPHASEs transversally is possible. We then
execute CPHASEs transversally between encoded logical qubits separated
by the $B$-$C$ wire interconnects.

 To achieve this is similar to the process described previously to reset the syndrome
bits and is illustrated in Fig.\ref{fig:fig2}(a). Instead of swapping the logical qubits onto the $C$-site, we instead simply create a redundant $z$-basis encoding across the $C$-site and the edge $A$-block qubit. We then transport each physical qubit of the encoded
logical qubits in each of the two interconnect-seperated $A$-blocks, via the $BC$-wires, onto the opposite $C$-sites, as described for the reset procedure. Once the two controlling qubits are located on the $C$-sites we execute a controlled $\frac{\pi}{8}$ phase gate between species $A$ and $C$, and subsequently
transfer them back to their respective $A$-blocks. One can show that this operation is equivalent to $\exp(-i\pi/4\,\sigma_z^{(i)}\sigma_z^{(i+1)})$, where $\sigma_z^{(i)},\,\sigma_z^{(i+1)}$, are operators on the nearest end physical qubits of interconnect-separated $A$-blocks. This, together with single qubit global operations on the $A$ and $C$ blocks (see Fig.\ref{fig:fig2}), yields a net result which is locally equivalent to a controlled-Z gate between the edge physical qubits between interconnect-seperated $A-$blocks. We now repeat this
for all physical qubits in the encoded logical qubits in the $A$-blocks to execute a
logical CPHASE between interconnect-separated $A$-blocks. This completes
the description of the base level quantum error correction step.
\begin{figure}[h]
\begin{centering}\setlength{\unitlength}{1cm} \begin{picture}(10,5)
\put(0,0){\includegraphics[width=9.5cm]{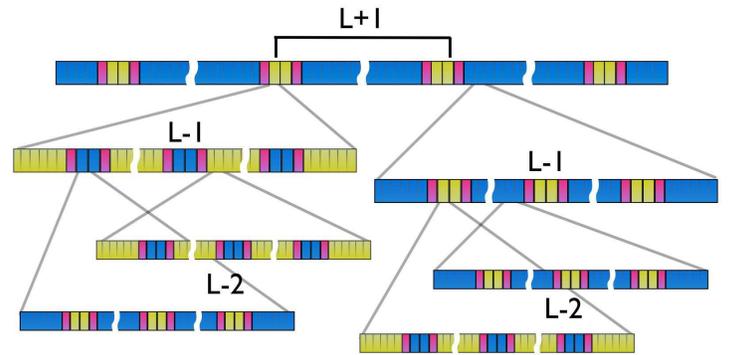}} 
 \end{picture} \par\end{centering}
\caption{Illustration of concatenation of levels $L+1,\cdots, L-2$. All $A$ (blue) and $B$ (yellow) cells are concatenated with self-similar encoding patterns while $C$-reset cells (pink) are not encoded.
\label{fig:fig3}}
\vspace{-.7cm}
\end{figure}

\textit{\bf Higher Concatenation Levels:-} To be useful we must devise
a method to concatenate the error correction in a manner which does
not require more species nor local addressing. Above we discussed
the level zero (single level encoding) concatenation, where we had
and $A-$blocks consisting of $N_{Comp}$ $A$-species cells linked together
by interconnects consisting $B$-wires and $C$-reset cells. In the level zero encoding discussed above, we have taken species $A$ to be the encoding species, storing the logical qubits, with the other species used to facilitate error correction and control over the species $A$ subchain. In what follows we will use $\tilde{A}_k$ to denote the $k^{th}$ level of encoding in this manner. We take $\tilde{A}_0$ to indicate a chain of $N_{Comp}$ species $A$ spins, $\tilde{A}_{0}=A^{\otimes N_{Comp}}$, and $\tilde{A}_{-1}$ to indicate two individual spins of species $A$. It will, however, also be necessary to consider regions where species $B$ holds the logical qubits, and species $A$ takes a facilitating role only. To this end, we will use $\tilde{B}_k$ to be the encoding achieved by swapping the roles of species $A$ and $B$ in $\tilde{A}_k$. Denoting
$\tilde{D}_{k}$ to now be the generalised $k^{th}$ concatenation
level interconnect, we set $\tilde{D}_{k}\equiv C \otimes \tilde{B}_{k-1}\otimes C$. 
In the base (or zero level), concatenation $\tilde{D}_{0}$ consists
of arrangements of $B$-wires and $C$-reset cells. We can consider the
level-1 meta-qubit {\em computational block}, $\tilde{A}_{1}$,
to consist of $N_{Comp}$ groups of the level-0 $A$-blocks and $\tilde{D}_{L_{0}}$
interconnects, $\tilde{A}_{1}=(\tilde{A}_{0}\otimes\tilde{D}_{0})^{\otimes N_{Comp}-1}\otimes\tilde{A}_{0}$. In general, we will take $\tilde{A}_{k}=(\tilde{A}_{k-1}\otimes\tilde{D}_{k-1})^{\otimes N_{Comp}-1}\otimes\tilde{A}_{k-1}$ for $k\ge1$, as illustrated in Fig.\ref{fig:fig3}. As the separation between the two unencoded species $C$ qubits in the interconnects, $\tilde{D}^{k-1}$, are two encoded qubits of increasing concatenation level, $\tilde{B}_{k-1}$, reset and controlled-$Z$ operations carried out on one level will not effect other levels of encoding. To make this more clear we show briefly how to engineer gates between species $A$ and $B$, and using these $AB$ gates, how to execute $Z$-rotations on meta-subchain end spins (required for universal quantum computation), CPHASE gates between interconnect-separated meta-$A$-blocks, and reset of the meta-ancillae.

\textit{$AB$-Gates:} We now build a $CZ_{A,B}$, a CPHASE gate between the $A$ and $B$ sites adjacent to a $C$ site. By noting that $CZ_{A,B} = CNOT_{A,C} CZ_{B,C} CNOT_{A,C} CZ_{B,C}$, where the $CNOT$'s target is given by the second index, and taking $CNOT_{A,C}=H_CCZ_{A,C}H_C$, and expressing $CZ_{A,C}=\exp(-i\pi/4 (\sigma_z^A+\sigma_z^C-\sigma_z^A\sigma_z^C))$, and gathering terms, one has
\begin{eqnarray}
CZ_{A,B} 
&= H_C  e^{i\frac{\pi}{4}(\sigma_z^{A} \sigma_z^{C})} R_z^{C}(-\frac{\pi}{4}) H_C R_z^{C}(-\frac{\pi}{4}) e^{i\frac{\pi}{4}(\sigma_z^{B} \sigma_z^{C})} H_C  \nonumber \\
& \times R_z^{C}(+\frac{\pi}{4}) e^{-i\frac{\pi}{4}(\sigma_z^{A} \sigma_z^{C})}
H_C   R_z^{C}(+\frac{\pi}{4}) e^{-i\frac{\pi}{4}(\sigma_z^{B} \sigma_z^{C})}
\end{eqnarray}
where $R_z^C(\theta)\equiv \exp(i\theta\sigma^C_z)$. This shows that the $CZ_{A,B}$ gate only requires local operations on the $C$-site and $A-C$, $C-B$, Ising interactions and with this we can perform transversal $CZ$ gates between the $A$ and $B$ species. \\
\textit{Rotations on $A$ meta-subchains:}must be level-$L$ dependent. This is achieved by conditioning their execution off the neighboring level-$L$ $B-$subchains. Through the pulse sequence
\begin{eqnarray}
R_z^{(ends)}(\theta) &=& CNOT_{B,A} X_B CNOT_{B,A} {R_z^A}(\frac{\theta}{2}) \nonumber \\
&\times&  CNOT_{B,A} X_B CNOT_{B,A} {R_z^A}(\frac{-\theta}{2}),
\end{eqnarray}
which uses the previous $CZ_{A,B}$ gate construction, we can effect a $Z-$rotation $R_z^A(\theta)$, on the end sites of the neighboring $A-$meta subchains. Those parts of the $A-$meta subchains not next to a $C$-site will experience $R_z^A(\theta/2)R_z^A(-\theta/2)$, the identity.\\
\textit{CZ Gates between $A-$meta subchains:}We again make use of the interconnecting level-$L$ meta-$B$-blocks to execute a CZ between the end sites of interconnect-separated level-$L$ $A$-subchains. Our construction will be such that the gate can be performed independently at any required concatenation level $L$. The gate is shown in Fig.\ref{fig:fig2}(b), and makes use of the $CNOT_{A,B}$ construction above and global rotations on the $A$-subchains. The latter cancel out for those parts of the $A$-subchains not next to a $C$ site. The level specific nature of the gate is embedded in the three mirror cycles of the level-$L$ $B-$subchain portion of the gate. To execute a CZ gate between two encoded meta-$A$-subchains one must perform CZ gates transversally on each element of the encoding.\\
\textit{Resetting the ancillae:}We have $C$-sites at the end of each level-$L$ encoded qubit. It is again vital that the ancillae reset occurs in a level specific manner as the qubits at other levels may be delocalised and must not be disturbed while we reset the ancillae at level-$L$. The circuit to achieve this again makes use of the triple level-$L$ meta-$B$-subchain mirroring and is shown in Fig.\ref{fig:fig2}(c), where we have used the above $CNOT_{A,C}$, etc. construction. When resetting encoded qubits, each element must brought to the ends of the level-$L$ $A-$ meta-subchain where they are then reset via the procedure in Fig.\ref{fig:fig2}(c).

\textit{\bf Proof of Threshold Existence:-} In order to prove the existence of a threshold for fault tolerant quantum computing within the system we will consider the error probability per gate at each level, $L$, of encoding, $P_L$. For a code which can correct one error per encoded qubit $P_L = \kappa P_{L-1}^2$ since all operations between qubits at level $L$ use only level $L-1$ operations (which have error probability $P_{L-1}$), and at least two errors are required to produce an error which is not correctable at the present level of encoding. We will take $N$ to be the number of level $L-1$ operations required to perform the level $L$ fault tolerant operation requiring the most level $L-1$ operations, plus one round of level $L$ error correction. Since the species $C$ chain never increases in length the effective error rate per physical qubit when doing controlled-$Z$ gates which cross the unencoded region is always constant, and bounded from above by 20 $\epsilon$ (the number of physical operations required to swap a qubit onto and then off a $C$-spin), where $\epsilon$ is the error probability per physical qubit per operation. This means that $N$, as defined above, is independent of concatenation level, $L$. As $\kappa$ is the number of ways in which an error uncorrectable at level $L-1$ can occur, it is strictly less than $\frac{N(N-1)}{2}$. This can probably be made smaller, but it suffices to show a threshold. Thus $P_L < \kappa^{2^L-1} \epsilon^{2^L}$. As $L$ goes to infinity, this limits to zero if epsilon is less than $1/\kappa$. Thus a threshold of $1/\kappa$ exists.

This work has been supported by the EC IST QAP Project Contract Number
015848. JF is supported by a Helmore Award.

\bibliography{QCAQEC}

\begin{thebibliography}{24}
\bibitem[{Foo()}]{Footnote1}
\bibinfo{note}{A small systematic error in the global pulse effectively yields
  independent Paul errors while an error on a $ZZ$ gate will yield at most a
  two correlated error event.}
\end{thebibliography}

\end{document}